\documentclass[twocolumn,showpacs]{revtex4}

\usepackage{graphicx}
\usepackage{dcolumn}
\usepackage{amsmath}
\usepackage{color}

\begin{document}

\title{Fluctuation-induced Nambu-Goldstone bosons in a Higgs-Josephson model
}

\author{Takashi Yanagisawa and Yasumoto Tanaka}

\affiliation{Electronics and Photonics Research Institute,
National Institute of Advanced Industrial Science and Technology (AIST),
Tsukuba Central 2, 1-1-1 Umezono, Tsukuba 305-8568, Japan
}


\begin{abstract}
We present a new mechanism of fluctuation-induced Nambu-Goldstone bosons in
a scalar field theory of Higgs-Josephson systems. 
We consider a simple scalar field model with $U(1)^n$ rotational symmetry.
When there is an interaction which violates the rotational symmetry, the
Nambu-Goldstone bosons become massive and massless bosons are concealed.
We present a model where
the massive boson becomes a massless boson as a result of the perturbative
fluctuation. In our model the ${\bf Z}_2$-symmetry associated with the chirality
is also broken.
The transition occurs as a weak first-order transition at the critical point.
The ground state at absolute zero will flow into the state with more
massless bosons due to fluctuation effects at finite temperature.
\end{abstract}


\maketitle

{\em Introduction}
When global and continuous symmetries are spontaneously broken,
gapless excitation modes, called the Nambu-Goldstone bosons, exist and
govern the long-distance behaviors of the system\cite{nam60,nam95,gol61}.
When the $U(1)$ rotational symmetry is spontaneously broken, there is
a massless Nambu-Goldstone boson.
When there is an interaction that violates the $U(1)$ symmetry,
we have no massless boson.
An interesting question is whether such an interaction will continuously
conceal the Nambu-Goldstone bosons when the perturbative corrections are taken
into account.
We present a model that exhibits a fluctuation induced Nambu-Goldstone boson
in this paper.
This means that a massless boson appears inspite of an interaction that
hides Nambu-Goldstone bosons.
We propose the mechanism of fluctuation induced Nambu-Goldstone boson.

We consider a model of an n-component scalar field with Josephson
interactions, so called the Higgs-Josephson model\cite{hig64,hig66,jos62,jos64}.
Let us consider the action given as
\begin{eqnarray}
S&=& \frac{1}{k_BT}\int d^dx\sum_j\left( \alpha_j|\phi_j|^2
+\frac{\beta_j}{2}|\phi_j|^4\right)\nonumber\\
&+& \frac{1}{k_B T}\int d^dx\Big[ \sum_jK_j|\nabla\phi_j|^2
+\sum_{i\neq j}\gamma_{ij}\phi_i^*\phi_j \Big], 
\end{eqnarray}
where 
$\phi\equiv(\phi_1,\cdots,\phi_n)$ is a complex n-component scalar field.
We write $\phi_j$ as
\begin{equation}
\phi_j= e^{i\theta_j}|\phi_j|=e^{i\theta_j}\rho_j,
\end{equation}
where $\rho_j$ ($j=1,\cdots,n$) are real scalar fields.
The last term in the action is the Josephson term.
We assume that $\gamma_{ij}$ are real and $\gamma_{ij}=\gamma_{ji}$.
Without this interaction, the phase modes $\theta_j$ $(j=1,\cdots,n)$ represent 
massless modes.
Because of this term, we have $n-1$ phase massive  modes and one
massless mode as shown by expanding
$\cos(\theta_i-\theta_j)$
in terms of $\theta_i-\theta_j$.
We adopt that $\beta_j$ is positive so that the action has a minimum.
When $\alpha_j$ is negative, $\rho_j$ takes a finite value at the 
minimum of the potential.  We set this value as $\Delta_j$ and
write $\rho_j=\Delta_j+H_j$.
$H_j$ is the Higgs field and represents fluctuation of the scalar field
around the minimum $\Delta_j$.
We simply assume that $K=K_j$, $\Delta=\Delta_j$ and
$\gamma_{ij}=\gamma_{ji}\equiv\gamma$.
Then the action for the phase variables $\theta_j$ is
\begin{equation}
S[\theta]= \frac{\Lambda^{d-2}}{t} \int d^dx\left( \sum_j(\nabla\theta_j)^2
+\lambda\Lambda^2\sum_{i< j}\cos(\theta_i-\theta_j) \right),
\end{equation}
where $t/\Lambda^{d-2}=k_BT/(K\Delta^2)$ and $\lambda\Lambda^2=\gamma/K$.
We have introduced the cutoff $\Lambda$ so that $t$ and $\lambda$ are
dimensionless parameters.
We assume that $\lambda>0$ in this paper.
We now focus on $\theta_j$ and consider the case $n=3$.
Since the potential term is written as
\begin{equation}
V\equiv (\lambda\Lambda^d/t)( \cos(\theta_1-\theta_2)+\cos(\theta_2-\theta_3)
+\cos(\theta_3-\theta_1) ),
\end{equation}
the mode of the total
phase $\theta_1+\theta_2+\theta_3$ remains massless.
We do not consider this mode because the coupling to the gauge field turns
this mode into a gapped mode (Higgs mechanism).
The other $n-1$ modes do not become massive by the coupling to the gauge field.
Let us consider the case $\lambda>0$.
As is easily shown, the ground state of $V$ has a $2\pi/3$ structure, namely,
$\theta_2-\theta_1=2\pi/3$ and $\theta_3-\theta_2=2\pi/3$ as shown in Fig.1(a).
The state in Fig.1(b) has also the same energy.  Two states are indexed by the
chirality $\kappa=1$ for (a) and $\kappa=-1$ for (b), where $\kappa$ is defined
by $\kappa= (2/3\sqrt{3})(\sin(\theta_2-\theta_1)+\sin(\theta_3-\theta_2)+
\sin(\theta_1-\theta_3)$\cite{miy84,oze03,has05,oku11,obu12,tan10a,tan10b,yan12}.
We set $\varphi_1=\theta_3-\theta_1$ and $\varphi_2=\theta_1-2\theta_2+\theta_3$
to write the potential density as
\begin{equation}
V= \frac{\lambda\Lambda^d}{t}\left( \cos(\varphi_1)
+2\cos\left(\frac{\varphi_1}{2}\right)\cos\left(\frac{\varphi_2}{2}\right)\right).
\end{equation}
$V$ has a minimum at $\varphi_1=4\pi/3$ and $\varphi_2=0$.
We mention here that an $S_3$ symmetry of the Josephson potential is not lost
when we express the potential in terms of $\varphi_1$ and $\varphi_2$.
When $V$ has a minimum at some value of $\varphi_1=\theta_3-\theta_1$,
$V$ has also a minimum when $\theta_3-\theta_2$ takes the same value
(modulo $2\pi$).  When the former has the chirality $\kappa=1$, the latter has 
$\kappa=-1$.
We consider the fluctuation around this minimum.  For this purpose, we perform
a unitary transformation by defining $\varphi_1=4\pi/3+\sqrt{2}\eta_1$ and
$\varphi_2=\sqrt{6}\eta_2$:
\begin{eqnarray}
\theta_1 &=&-\frac{2\pi}{3}-\frac{1}{\sqrt{2}}\eta_1+\frac{1}{\sqrt{6}}\eta_2
+\frac{1}{\sqrt{3}}\eta_3,\\
\theta_2 &=& ~~~~~~~~-\frac{2}{\sqrt{6}}\eta_2+\frac{1}{\sqrt{3}}\eta_3,\\
\theta_3 &=& \frac{2\pi}{3}+\frac{1}{\sqrt{2}}\eta_1+\frac{1}{\sqrt{6}}\eta_2
+\frac{2}{\sqrt{3}}\eta_3.
\end{eqnarray}
where $\eta_i$ ($i=1,2,3$) indicate fluctuation fields.
$\eta_3$ describes the total phase mode, 
$\eta_3=(\theta_1+\theta_2+\theta_3)/\sqrt{3}$, and is not important in this paper
because this mode turns out to be a plasma mode by coupling with the
long-range Coulomb potential.
We obtain $\sum_i(\nabla\theta_i)^2=\sum_i(\nabla\eta_i)^2$, and then the action
$S[\eta]\equiv S[\theta]$ is
\begin{eqnarray}
S[\eta]&=& \frac{\Lambda^{d-2}}{t}\int d^dx\Big[ \sum_j(\nabla\eta_j)^2
+\lambda\Lambda^2\Bigl( \cos\left( \sqrt{2}\eta_1+\frac{4\pi}{3}\right)\nonumber\\
&+& 2\cos\left( \frac{1}{\sqrt{2}}\eta_1+\frac{2\pi}{3}\right)
\cos\left( \sqrt{\frac{3}{2}}\eta_2\right) \Bigr) \Big].
\end{eqnarray}
The potential term has a minimum at $\eta_1=\eta_2=0$.
Both of $\eta_1$ and $\eta_2$ represent massive modes with mass
$3\lambda/(2t)$.

{\em Fluctuation induced Nambu-Goldstone boson}
The potential $V$ corresponds to the potential of a two-dimensional XY model
on the triangular lattice with a frustrated interaction.
The ground state has an well known $2\pi/3$-structure.
We consider the role of fluctuation and show the existence of 
fluctuation-induced massless mode.
We examine the following free-energy density by neglecting the kinetic term:
\begin{eqnarray}
f&=& k_BT\frac{\lambda\Lambda^d}{t}\Big[ \cos\left( \sqrt{2}\eta_1
+\frac{4\pi}{3}\right)\nonumber\\ 
&+& 2\cos\left( \frac{1}{\sqrt{2}}\eta_1+\frac{2\pi}{3}\right)
 \cos\left( \sqrt{\frac{3}{2}}\eta_2\right) \Big].
\end{eqnarray}
The partition function is given by
\begin{equation}
Z= \int d\eta_1 d\eta_2\exp\left( -\frac{F}{k_BT}\right),
\end{equation}
for the free energy functional $F$.
Using the formula for the modified Bessel function,
\begin{equation}
I_0(z)= \frac{1}{\pi}\int_0^{\pi}e^{z\cos\varphi}d\varphi,
\end{equation}
we have, by using $\varphi_2/2=\sqrt{3/2}\eta_2$,
\begin{eqnarray}
&&\int_0^{2\pi}d\varphi_2 \exp\Big[ -\frac{2\lambda\Lambda^d}{t}
\cos\left(\frac{\varphi_1}{2}\right)
\cos\left(\frac{\varphi_2}{2}\right) \Big] \nonumber\\
&& ~~~~~= 2\pi I_0\left( \frac{2\lambda\Lambda^d}{t}\cos\left(\frac{\varphi_1}{2}
\right) \right).
\end{eqnarray} 
We use $I_0(-x)=I_0(x)$ and the asymptotic formula 
$I_0(z)\sim e^z/\sqrt{2\pi z}$ ($z>0$) at low temperature.
Then the effective free-energy density for $\eta_1$ is
\begin{eqnarray}
\frac{f[\eta_1]}{\Lambda^d}&=& \epsilon_0\cos\left(\sqrt{2}\eta_1+\frac{4\pi}{3}
\right)
-2\epsilon_0\Bigl|\cos\left(\frac{1}{\sqrt{2}}\eta_1+\frac{2\pi}{3}\right)
\Bigr|\nonumber\\
&+& \frac{1}{2}\frac{k_BT}{\Lambda^d}
\ln\left( \frac{\lambda\Lambda^d}{\pi t}\Bigl| 
\cos\left( \frac{1}{\sqrt{2}}\eta_1+\frac{2\pi}{3}\right)\Bigr|\right),
\end{eqnarray}
where $\epsilon_0\equiv k_BT\lambda/t$.
We have an effective entropy term being proportional to the temperature $T$.
$F[\eta_1]$ has a minimum at $\eta_1=0$ ($\varphi_1=4\pi/3$) at absolute zero 
$T=0$.
In contrast, at finite temperature $T>0$, the minimum is at
$\eta_1=-\sqrt{2}\pi/6$ and $\varphi_1=\pi$.
This is shown in Fig.2 where we present the potential $f[\eta_1]/\epsilon_0$ 
as a function of
$\varphi\equiv \varphi_1=4\pi/3+\sqrt{2}\eta_1$ for $t=\lambda$ with
setting $\Lambda=1$.
At $\varphi=\pi$, $\eta_2$ becomes a massless boson because the
free-energy density in eq.(10) becomes independent of $\eta_2$ with
vanishing of the mass term. 
This is due to the fluctuation of $\eta_2$ field at finite temperature.
The qualitatively same result is obtained by the Gaussian integration with 
respect to $\eta_2$
after expanding cosine function as $\cos(\sqrt{3/2}\eta_2)=1-(3/4)\eta_2^2+\cdots$
and assuming that $\cos(\eta_1/\sqrt{2}+2\pi/3)<0$.
(We can use the formula $I_0(z)\sim e^z/\sqrt{2\pi z}$ when $z>0$ is large.
In the limit $z\rightarrow 0$, we have a largee entropy  coming from the
volume of the phase space and thus the minimum is at $\varphi_1=\pi$ for $T>0$
when $T$ is higher than a critical value.)
This state is shown in Fig.1(c) using a spin analogue where we have two
antiferromagnetic spins and one vanishing spin.
This means that the $\eta_2$-mode is massless and
$\varphi_2=\sqrt{6}\eta_2=\theta_1-2\theta_2+\theta_3$ can take any value.
At the absolute zero, we have the index of chirality $\kappa=\pm 1$ as
shown in Figs.1(a) and 1(b).
The chirality disappears at finite temperature leading to the
emergency of a Nambu-Goldstone boson.
This represents a phenomenon that the Nambu-Goldstone boson appears due to 
a fluctuation effect.

\begin{figure}[htbp]
\begin{center}
\includegraphics[height=7.5cm,angle=90]{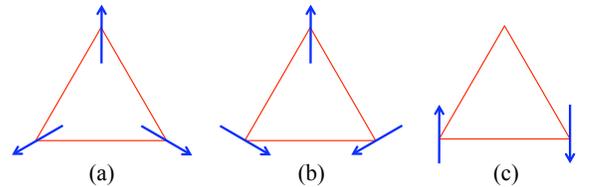}
\caption{
(a) $2\pi/3$-structure in the ground state with the chirality $\kappa=1$.
(b) Degenerate ground state with the chirality $\kappa=-1$.
(c) Spin structure at finite temperature.  Two spins are antiferromagnetically
aligned and one spin vanishes, that is, the expectation value vanishes: 
$\langle{\bf S}\rangle=0$.
This means that the one spin is rotating freely, indicating the existence of a
massless boson.
}
\end{center}
\end{figure}

\begin{figure}[htbp]
\begin{center}
\includegraphics[width=7cm]{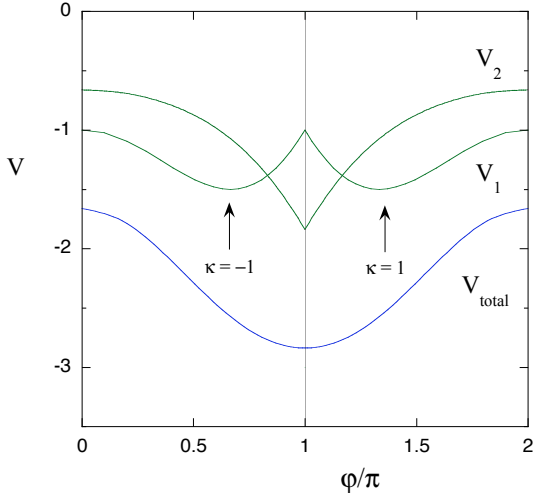}
\caption{
Potential terms $V_1$, $V_2$ and $V_{total}=V_1+V_2$ as a function of
$\varphi\equiv 4\pi/3+\sqrt{2}\eta_2$ for $t/\lambda=1$ and $\Lambda=1$.
$V_1=\cos(\varphi)-2|\cos(\varphi/2)|$ and 
$V_2=2|\cos(\varphi/2)|-(t/\lambda)\ln(2\pi I_0(2(\lambda/t)|\cos(\varphi/2)|))$.
$V_1$ and $V_2$ are symmetric with respect to the axis of $\varphi=\pi$.
The aymptotic form of $\epsilon_0(V_1+V_2)$ agrees with eq.(14).
The total potential $V_{total}=V_1+V_2$ has a minimum at $\varphi=\pi$
due to the logarithmic term.
Minima of $V_1$ correspond to the state of chirality $\kappa=1$ and
$\kappa=-1$, respectively.
}
\end{center}
\end{figure}

{\em Phase transition at finite temperature}
We next consider the kinetic terms of $\eta_j$.  For this purpose, we use
the expansion of cosine term 
and write the action
in the form
\begin{eqnarray}
S&=& \frac{\Lambda^{d-2}}{t}\int d^dx\Big[ \sum_j(\nabla\eta_j)^2
+\lambda\Lambda^2\Bigl( \cos\left(\sqrt{2}\eta_1+\frac{4\pi}{3}\right)\nonumber\\
&-& 2\Bigl|\cos\left( \frac{1}{\sqrt{2}}\eta_1+\frac{2\pi}{3}\right)\Bigr|\Bigr)
+\frac{3\lambda\Lambda^2}{2}\Bigl|\cos\left( \frac{1}{\sqrt{2}}\eta_1+
\frac{2\pi}{3}\right)\Bigr|\eta_2^2 \Big]. \nonumber\\
\end{eqnarray}
When $\cos(\varphi_1/2)<0$, we use $\cos(\sqrt{3/2}\eta_2)=1-(4/3)\eta_2^2+\cdots$. 
Around the minimum at $\varphi_1=2\pi/3$ and
$\varphi_2=2\pi$ (with chirality $\kappa=-1$), we use instead the expansion by
defining $\varphi_2=2\pi+\sqrt{6}\eta_2$.
We integrate out the field $\eta_2$ to obtain the effective action, 
using $\varphi\equiv\varphi_1=4\pi/3+\sqrt{2}\eta_1$,
\begin{eqnarray}
S&=& \frac{\Lambda^{d-2}}{t}\int d^dx\Big[ \frac{1}{2}(\nabla\varphi)^2
+\lambda\Lambda^2\Bigl( \cos\varphi
-2\Bigl|\cos\left(\frac{\varphi}{2}\right)\Bigr|\Bigr)\Big]\nonumber\\
&+& \frac{1}{2}Tr\ln\left( -\frac{\Lambda^{d-2}}{t}\nabla^2+\frac{3\lambda}{2t}
\Lambda^2
\Bigl| \cos\left(\frac{\varphi}{2}\right)\Bigr|\right).
\end{eqnarray}
When we neglect the kinetic term $-\nabla^2$, this action is reduced to
the previous effective free energy.
We adopt that the spatial variation of $\varphi$ field is very slow so that
we can perform the ${\bf k}$-summation for $-\nabla^2={\bf k}^2$.
In the two-dimensional case $(d=2)$, the effective free-energy density is
obtained as
\begin{eqnarray}
\frac{f[\varphi]}{\Lambda^2}&=&  \frac{1}{2}K\Delta^2\Lambda^{-2}(\nabla\varphi)^2
+\epsilon_0\Bigl( \cos\varphi
-2\Bigl|\cos\left(\frac{\varphi}{2}\right)\Bigr|\Bigr)\nonumber\\
&+& \frac{1}{2}k_BT\frac{c}{4\pi}\ln\left(\frac{c\Lambda^d}{t}
+\frac{3\lambda\Lambda^d}{2t}
\Bigl|\cos\left(\frac{\varphi}{2}\right)\Bigr|\right)\nonumber\\
&+& k_BT\frac{3\lambda}{16\pi}
\Bigl| \cos\left(\frac{\varphi}{2}\right)\Bigr|
\ln\Bigl(1+\frac{2c}{3\lambda}
\Bigl| \cos\left(\frac{\varphi}{2}\right)\Bigr|^{-1}\Bigr),
\nonumber\\
\end{eqnarray}
where we have chosen the cutoff $k_0$ in the momentum space as
$k_0^2=c\Lambda^2$ for a constant $c$.

\begin{figure}[htbp]
\begin{center}
\includegraphics[width=7cm]{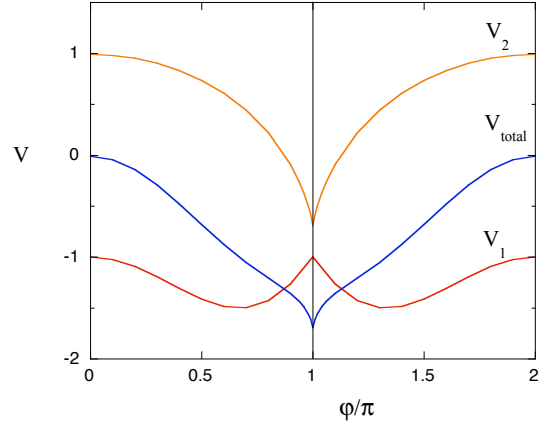}
\caption{
Potential terms $V_1$, $V_2$ and $V_{total}=V_1+V_2$ as a function of
$\varphi\equiv 4\pi/3+\sqrt{2}\eta_2$ for $t=8\pi$ and $\lambda/c=1$ where
we set $c=4\pi$ to compare with $V$ in Fig.2.
$V_1$ is the same as that in Fig.2 and $V_2$ is 
$V_2=(t/2\lambda)(c/4\pi)\ln(c/t+(3\lambda/2t)|\cos(\varphi/2)|)
+(3t/16\pi)|\cos(\varphi/2)|
\ln(1+(2c/3\lambda)|\cos(\varphi/2)|^{-1})$.
The total potential $V_{total}=V_1+V_2$ has a minimum at $\varphi=\pi$.
}
\end{center}
\end{figure}

\begin{figure}[htbp]
\begin{center}
\includegraphics[width=7cm]{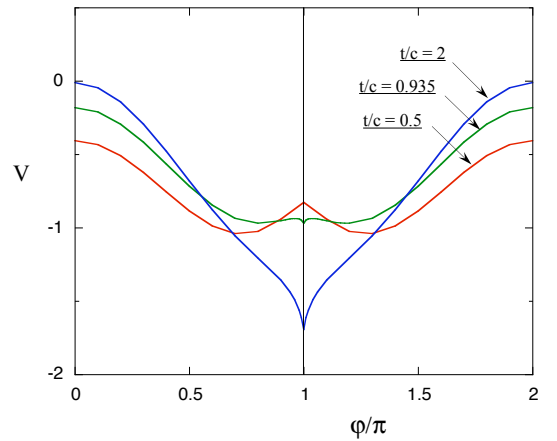}
\caption{
Potential as a function of $\varphi$ for $t/c=0.5$, $0.935$ and $2$, 
respectively,
where we set $\lambda/c=1$ and $c=4\pi$.
}
\end{center}
\end{figure}

The spatial fluctuation softens the thermal fluctuation effect and there
is a finite critical temperature where the minimum at $\varphi=4\pi/3$
disappears and simultaneously the chirality vanishes.
We show the potential term as a function of $\varphi$ for $t=2c$
and $\lambda=c$ with $c=4\pi$ in Fig.3 where we subtracted the term 
$k_BT/2\ln\Lambda^d$
which is independent of $\varphi$ (or equivalently we set $\Lambda=1$).  
We have a minimum at $\varphi=\pi$ when $t$ is large as shown in Fig.3.
The critical temperature $t_c$ is scaled by $\lambda/c$:
\begin{equation}
t_c = t_c(\lambda/c).
\end{equation}
$t_c$ is estimated by the equation
$V(\varphi=4\pi/3)=V(\varphi=\pi)$, which gives
\begin{equation}
\frac{k_BT_c}{K\Delta^2}=t_c=\frac{\lambda}{\frac{c}{4\pi}
\ln\left(1+\frac{3\lambda}{4c}\right)
+\frac{3\lambda}{16\pi}\ln\left(1+\frac{4c}{3\lambda}\right)}.
\end{equation}
For small $\lambda\rightarrow 0$, $t_c$ is small:
$t_c\simeq 16\pi/(3\ln(1/3\lambda))$.  When $\lambda$ is large, $\lambda\gg 1$,
$t_c$ is also large $t_c\simeq 4\pi\lambda/c\ln(3\lambda/4c)$.
In Fig.4 we show the potential for $t/c=0.5$, $0.935$,
$2$, and $\lambda/c=1$ with $c=4\pi$.
When $t$ is small, the potential has a minimum at $\varphi=4\pi/3$ or
at $\varphi=2\pi/3$ indicating
that the ground state has the $2\pi/3$ structure with the chirality $\pm 1$.
In contrast, when $t$ is large, we have a minimum $\varphi=\pi$.
There is a transition at finite temperature $t=t_c$.
This is a first-order transition since we have the double-minimum potential
in the range $\pi\le\varphi\le 2\pi$.
This should be called a weak first-order transition because the change of
$V_{total}(\varphi=\pi)$ is slow as $t$ is varied near the critical 
temperature.
The minimum point of $\varphi$ changes gradually from $4\pi/3$ and
changes suddenly to $\pi$ at the critical temperature.
For $t>t_c$, the $\eta_2$-mode represents a massless boson.
We show $t_c$ as a function of $\lambda/c$ in Fig.5.

\begin{figure}[htbp]
\begin{center}
\includegraphics[width=7cm]{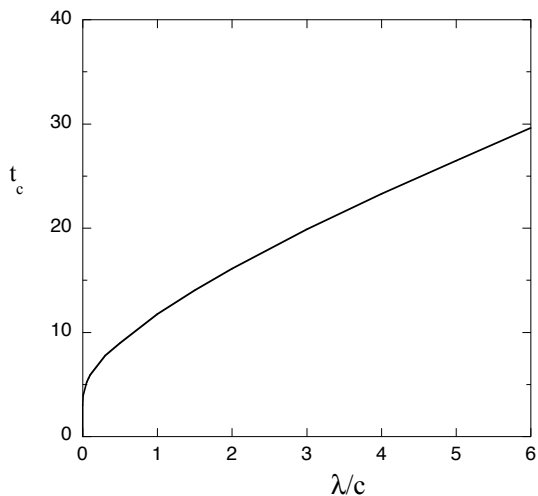}
\caption{
$t_c$ as a function of $\lambda/c$. 
where we set $\lambda/c=1$ and $c=4\pi$.
$t_c$ is a increasing function of $\lambda/c$.
}
\end{center}
\end{figure}

We discuss a relation to the classical XY model on a two-dimensional triangular 
lattice.
The ground state of the 2D XY model has the $2\pi/3$-structure to minimize 
the energy.  There is a transition of the chirality at finite temperature.
The critical temperature $T_c$ is of the order of the exchange coupling $J$
because $\lambda/t=J/k_BT$ in this case.
The Kosterlitz-Thouless (KT) transition also occurs in the XY model on
the 2D triangular lattice.
The critical temperature of the KT transition $T_{KT}$ is determined by
the renormalization group equation.  In general, $T_{KT}$ is different
from the critical temperature of the chiral transition
$T_{chiral}\equiv K\Delta^2 t_c$.

\begin{figure}[htbp]
\begin{center}
\includegraphics[width=7cm]{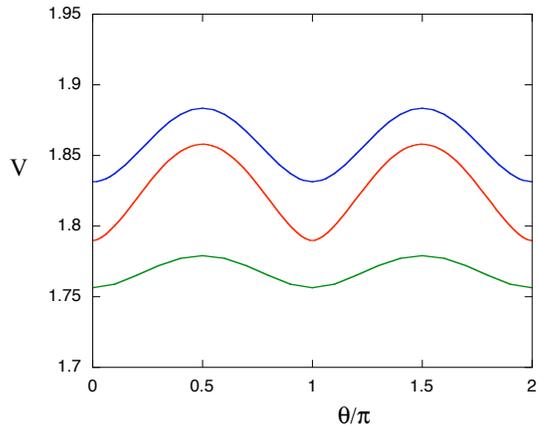}
\caption{
Effective potential as a function of $\theta$ for $n=4$. 
From the top, we set $\lambda=1$ and $a=1.2$, $\lambda=1$ and $a=1.001$ and
$\lambda=0.5$ and $a=1.2$, respectively,
The potential has mimima at $\theta=m\pi$ for integer $m$.
}
\end{center}
\end{figure}

The similar phenomenon occurs for an $n=4$ theory with the potential
\begin{eqnarray}
V&=& \frac{\lambda\Lambda^d}{t}\Big[ \cos(\theta_1-\theta_2)+a\cos(\theta_1-\theta_3)
+\cos(\theta_1-\theta_4)\nonumber\\
&+& \cos(\theta_2-\theta_3)+a\cos(\theta_2-\theta_4)+\cos(\theta_3-\theta_4)\Big],
\end{eqnarray}
where $a\ge 1$ is a constant.
This model has a close relation with the 2D antiferromagnetic XY model
on a square lattice\cite{hen89,loi00}.
One of the ground state is given by 
$(\theta_1,\theta_2,\theta_3,\theta_4)=(0,\theta,\pi,\theta+\pi)$ where real
$\theta$ is arbitrary and the ground state is degenerate with respect to
$\theta$.
We define $\varphi_1=\theta_1-\theta_2-\theta_3+\theta_4=\eta_1$,
$\varphi_2=\theta_1+\theta_2-\theta_3-\theta_4=\eta_2-2\pi$,
$\varphi_3=\theta_1-\theta_2+\theta_3-\theta_4=\eta_3-2\theta$, and
the total phase $\Phi=\theta_1+\theta_2+\theta_3+\theta_4$.
Then the potential becomes
\begin{equation}
V= \frac{\lambda\Lambda^d}{t}\Big[ -2a+\frac{1}{4}(a-\cos\theta)\eta_1^2
+\frac{1}{4}(a+\cos\theta)\eta_2^2+\cdots \Big],
\end{equation}
where $\cdots$ indicates higher order terms. 
The $\eta_3$-mode becomes massless and the ground state energy $-2a$ is
independent of $\theta$.  This is the $n-3$ series state\cite{yan13} 
which we call the type I.
When $a=1$, $\eta_1$- or $\eta_2$-mode is massless in the case $\theta=0$ or $\pi$.
This is the $n-2$ series state.
The effective potential $V_{eff}$ is obtained by integrating out the $\eta_1$
and $\eta_2$ variables in a similar way to the case $n=3$ in two dimensions:  
\begin{eqnarray}
\frac{V_{eff}}{k_B T\Lambda^2}&=& \frac{1}{2}\ln\left( (4\pi+\lambda a)^2
-\lambda^2\cos^2\theta \right)\nonumber\\
&+&\frac{1}{8\pi}\lambda a\ln\left( \frac{(4\pi+\lambda a)^2-\lambda^2\cos^2\theta}
{\lambda^2(a^2-\cos^2\theta)}\right)\nonumber\\
&+& \frac{\cos\theta}{8\pi}\ln\left( \frac{4\pi+\lambda(a+\cos\theta)}
{\lambda(a+\cos\theta)}\frac{\lambda(a-\cos\theta)}{4\pi+\lambda(a-\cos\theta)}
\right),\nonumber\\
\end{eqnarray} 
where we used the cutoff $k_0$ in the momentum integral satisfying
$k_0^2/(4\pi)=\Lambda^2$. 
The potential is shown in Fig.6 for several parameters where
the ground state is at $\theta=m\pi$ for an integer $m$. 
This indicates that a Nambu-Goldstone
boson emerges for $a=1$ as a result of fluctuation of the $U(1)$ phase variables.
We can regard the sign of $\sin\theta$ as a kind of chirality.
The emergence of new massless boson is accompanied by the vanishing of
chirality.
 
We can generalize our argument to an $n$-component scalar field with
Josephson couplings.
The potential
\begin{equation}
V= \frac{\lambda\Lambda^d}{t}\sum_{i<j}\cos(\theta_i-\theta_j),
\end{equation}
has a series of massless bosons; there are two types of ground states called
the type I and II\cite{yan13}.  
In the ground state I one has $n-3$ massless bosons and
in the ground state II one has $n-2$ massless bosons.
(The $n-2$ series exists only for even $n$.)
Two ground states I and II are degenerate for the potential $V$.
However, the ground state II becomes more stable than the state I due to
fluctuation effect.
Thus, when we are in the ground state I first, the fluctuation effect
leads us to the state II with increasing the number of Nambu-Goldstone bosons.

{\em Order to order transition by disorder}
The chiral transition considered in this paper is a transition from
the $2\pi/3$-structure in Fig.1(a) (or (b)) to the antiferromagnetic
state in Fig.1(c).  We can say that the ordered state with a massless
boson in Fig.1(c) is induced by disorder, namely, thermal fluctuation.
We call this an order to order transition by disorder.
We discuss here the fluctuation effect on the induced Nambu-Goldstone
boson.
For this purpose, we write $\varphi_1=\pi+\phi_1$ so that $\phi_1$ indicates
the fluctuation mode in the neighborhood of $\pi$.  The action is written
as
\begin{eqnarray}
S&=& \frac{\Lambda^{d-2}}{t}\int d^d x\Big[ \frac{1}{2}(\nabla\phi_1)^2
+\frac{1}{2}\lambda\Lambda^2\phi_1^2+\frac{1}{6}(\nabla\varphi_2)^2\nonumber\\
&& ~~~~-\lambda\Lambda^2\phi_1\cos\left(\frac{\varphi_2}{2}\right)\Big].
\end{eqnarray}
The $\varphi_2$-mode is obviously a massless mode, but there is an
interaction with $\phi_1$.  This interaction will generate an effective 
potential of $\varphi_2$ that is proportional to 
$\cos^2(\varphi_2/2)=(\cos(\varphi_2)+1)/2$.  Then, the effective action for
$\varphi_2$ is given by the sine-Gordon model:
\begin{equation}
S_{\varphi_2}= \frac{\Lambda^{d-2}}{t}\int d^d x\Big[
\frac{1}{6}(\nabla\varphi_2)^2-\frac{\lambda}{4}\Lambda^2
\cos(\varphi_2)\Big].
\end{equation}
The low-energy property is determined by the values of $\lambda$ and $t$
as indicated by the renormalization group
equations\cite{ami80,zinn} near two dimensions.
The critical value of $t$, denoted by $t_{2c}$, is $t_{2c}=8\pi/3$.
We assume that $t>t_c>t_{2c}$.
When $\lambda$ is small, $\lambda$ is renormalized to $0$ following
the renormalization flow. This indicates
that the $\varphi_2$-mode remains massless for small $\lambda$.
When $\lambda$ is large, $\lambda$ is renormalized to be a large value,
showing that the potential term dominates the behavior of $\varphi_2$-mode 
and then that $\varphi_2$ takes the value near $0$.
In this case the massless $\varphi_2$-mode becomes massive, that is,
a gapped mode again.
Basically $\varphi_2$-mode may remain massless because
the Josephson coupling $\lambda$ is small in real superconductors.

{\em Summary}
We have proposed the mechanism of fluctuation induced Nambu-Goldstone bosons.
In an n-component scalar field theory with frustrated Josephson interactions,
massless bosons appear due to fluctuations at finite temperature.
In the 3-component theory discussed in the paper, a massless boson
appears and the chirality vanishes as the temperature is increased, that is,
the $Z_2$-symmetry breaking is driven by the chirality.
This shows that nature prefers massless bosons.  In fact,
in an n-component model, the ground state at absolute zero will flow into
the state with more massless bosons as the temperature is increased from
$(n-3)$-state to $(n-2)$-state.

The excitation modes in our model has an analogy to the vibration modes
of a molecule CH$_2$.  
Two modes, the scissoring mode and the rocking mode, are important
in determining the excitation spectra of CH$_2$\cite{sea82,leo84}.
The modes shown by $\varphi_1=\theta_3-\theta_1$ and 
$\varphi_2=\theta_1-2\theta_2+\theta_3$
represent the scissoring and rocking modes, respectively.
In our model, the rocking mode plays a significant role.
The fluctuation effect of the rocking mode becomes large as the temperature is
increased and gives rise to the phase transition.
The model presented in the paper appears as an effective free energy in 
multi-band superconductors\cite{tan10a,tan10b,yan12,sta10,dia11,lin12,wil13,mal13}.
Low energy excitation states are important in superconductors, the existence
of massless modes have been pointed out\cite{yan13,lin12,kob13,tan11}.
In this paper we have presented a new mechanism of the emergence of
Nambu-Goldstone bosons.

We express our sincere thanks to J. Kondo and I. Hase for useful discussions.
This work was supported by a Grant-in-Aid for Scientific Research from the
Ministry of Education, Culture, Sports, Science and Technology of Japan.

\end{document}